\input{aipcheck}

\documentclass[
    ,draft            
  ]
  {aipproc}

\layoutstyle{6x9}

\begin{document}
\newcommand{\etal}{{\sl et al. }}
\newcommand{\degs}{$^{\circ}$}
\newcommand{\ergs}{$\,$erg$\,$s$^{-1}$}
\newcommand{\eg}{{\sl e.g. }}

\title{Exploring the Nature of Weak Chandra Sources near the Galactic Centre}


\author{R. M. Bandyopadhyay}{
  address={Oxford University}
}

\author{J. C. A. Miller-Jones}{
  address={Oxford University}  
 ,altaddress={University of Amsterdam} 
}

\author{K. M. Blundell}{
  address={Oxford University}}

\author{F. E. Bauer}{
  address={Institute of Astronomy, Cambridge University}
  ,altaddress={Columbia University} 
}

\author{Ph. Podsiadlowski}{
  address={Oxford University}}

\author{Q. D. Wang}{
  address={University of Massachusetts}
}

\author{S. Rappaport}{
  address={MIT}
}

\author{E. Pfahl}{
  address={University of Virginia}
}

\begin{abstract}
We present results from the first near-IR imaging of the weak X-ray
sources discovered in the {\it Chandra}/ACIS-I survey (Wang \etal
2002) towards the Galactic Centre (GC). These $\sim$800 discrete
sources, which contribute significantly to the GC X-ray emission,
represent an important and previously unknown population within the
Galaxy. From our VLT observations we will identify likely IR
counterparts to a sample of the hardest sources, which are most likely
X-ray binaries. With these data we can place constraints on the nature
of the discrete weak X-ray source population of the GC. 
\end{abstract}

\maketitle


\section{The {\it Chandra} Galactic Centre Survey}

In July 2001 Wang \etal (2002) performed an imaging survey with {\it
Chandra}/ACIS-I of the central 0.8$\times$2\degs of the Galactic
Centre (GC), revealing a large population of previously undiscovered
discrete weak sources with X-ray luminosities of
$10^{32}-10^{35}$\ergs.  The nature of these $\sim$800 newly detected
sources, which may contribute $\sim$10\% of the total X-ray emission of
the GC, is as yet unknown.  In contrast to the populations of faint
AGN discovered from recent deep X-ray imaging out of the Galactic
plane, our calculations suggest that the extragalactic contribution to
the hard point source population over the entire Wang \etal survey is
$\leq$ 10\%, consistent with the log(N)-log(S) function derived from
the {\it Chandra} Deep Field data (\eg Brandt \etal 2001).  The harder
($\geq$3 keV) X-ray sources (for which the softer X-rays have been
absorbed by the interstellar medium) are likely to be at the distance
of the GC, while the softer sources are likely to be foreground X-ray
active stars or cataclysmic variables (CVs) within a few kpc of the
Sun.  The distribution of X-ray colours (Figure 1) suggests that only
a small fraction of the {\it Chandra} sources are foreground objects.
The combined spectrum of the discrete sources shows emission lines
characteristic of accreting systems such as CVs and X-ray binaries
(XRBs).  These hard, weak X-ray sources in the GC are therefore most
likely a population of XRBs; candidate classes include quiescent black
hole binaries or quiescent low-mass XRBs, CVs, and high-mass
wind-accreting neutron star binaries (WNSs).

\begin{figure}
  \includegraphics[width=.98\textwidth]{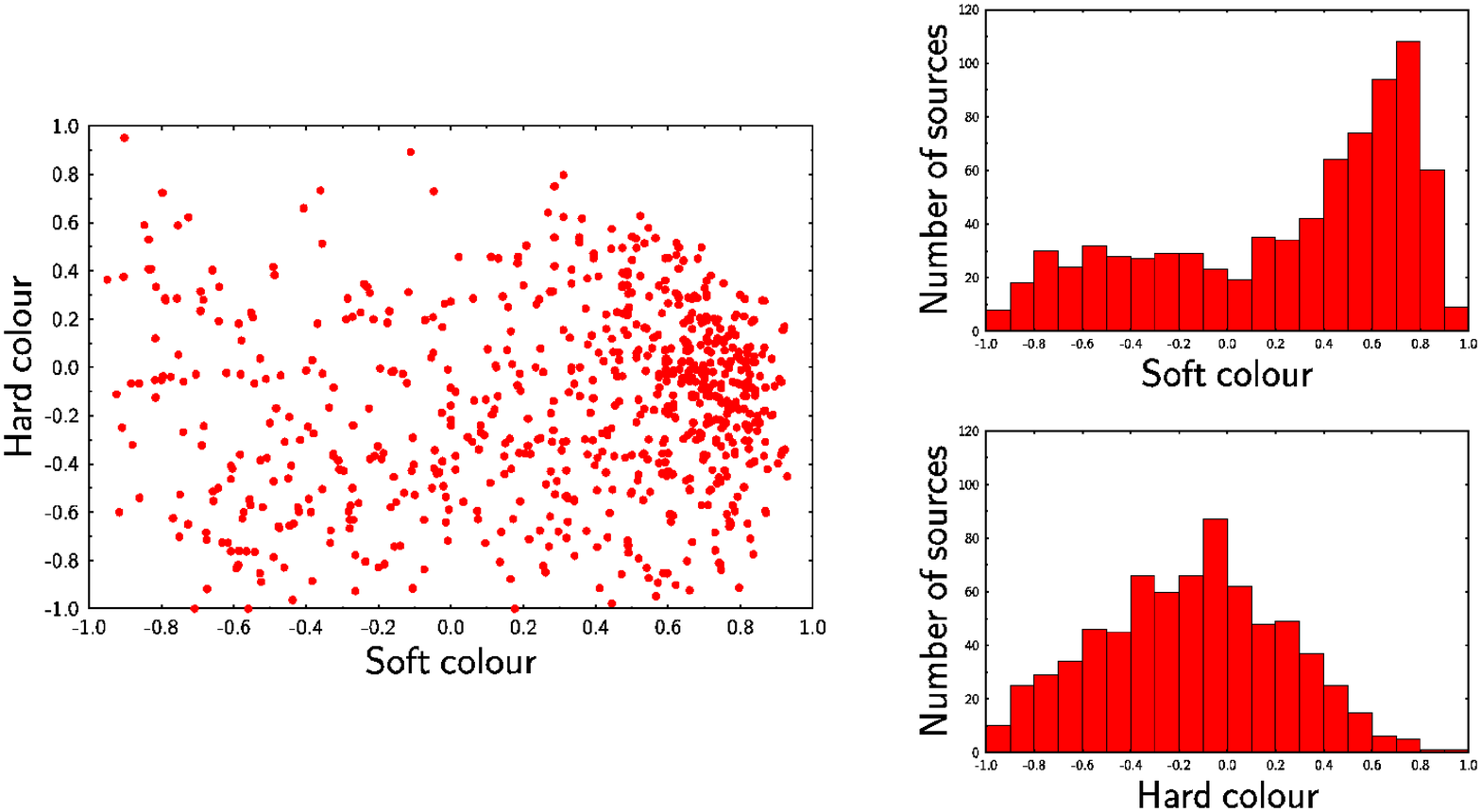}
  \caption{Characteristics of the X-ray source population detected in the {\it Chandra} mosaic.  Left panel: colour-colour diagram; right panel: histogram of the number of soft and hard X-ray sources in the GC field.  Soft colour is defined using the 1-3 and 3-5 keV bands, and hard colour using the 3-5 and 5-8 keV bands.  The distribution of X-ray colours suggests that only a small fraction of the {\it Chandra} sources are foreground objects.}
\end{figure}

\section{What are these Point Sources?}

Pfahl \etal (2002) have considered in detail the likely nature of
these {\it Chandra} sources and concluded on the basis of binary
population synthesis (BPS) models that many, if not the majority, of
these systems are WNSs. Depending on the mass of the companions, the
WNSs may belong to the ``missing'' population of wind-accreting
Be/X-ray transients in quiescence or the progenitors of
intermediate-mass X-ray binaries (IMXBs; 3$\leq
M$/\mbox{$M_{\odot}$}$\leq$7).  The existence of tens of thousands of
quiescent Be/XRBs in the Galaxy has been predicted since the early
1980s (Rappaport \& van den Heuvel 1982; Meurs \& van den Heuvel
1989), while it has only recently been recognized that IMXBs may
constitute a very important class of XRBs that had not been considered
before (King \& Ritter 1999; Podsiadlowski \& Rappaport 2000).  The
Wang \etal {\it Chandra} survey may contain as many as 10\% of the
entire Galactic population of WNSs.  In addition to the WNSs, Pfahl
\etal estimate that a small fraction of the {\it Chandra} sources
could be CVs or transient low-mass XRBs/black-hole binaries.

\section{Our VLT Imaging Program}

The first step in determining the nature of this population is to
identify counterparts to the X-ray sources.  The successful
achievement of our goals requires astrometric accuracy and high
angular resolution to overcome the confusion limit of the crowded GC.
The 2MASS survey is severely confusion limited in the GC and is of
insufficient depth ($K$=14.3) to detect the majority of the expected
counterparts.  We therefore constructed a survey program using the
ISAAC IR camera on the VLT to obtain high-resolution $JHK$ images in
order to identify a statistically significant number of counterparts
to the X-ray sources on the basis of the {\it Chandra} astrometry.  We
imaged 26 fields within the {\it Chandra} survey region, containing a
total of 85 X-ray sources.  In constructing our VLT program, we
preferentially selected for hard X-ray sources from the {\it Chandra}
survey, as the soft sources are most likely to be foreground.  Of the
hardest sources we selected those detected with an S/N$\geq$3 and
which were imaged in the central area of the ACIS-I field (due to the
off-axis characteristics of the {\it Chandra} PSF, greater astrometric
accuracy can be obtained for sources in the central regions of the
field).

\begin{figure}
  \includegraphics[width=.5\textwidth]{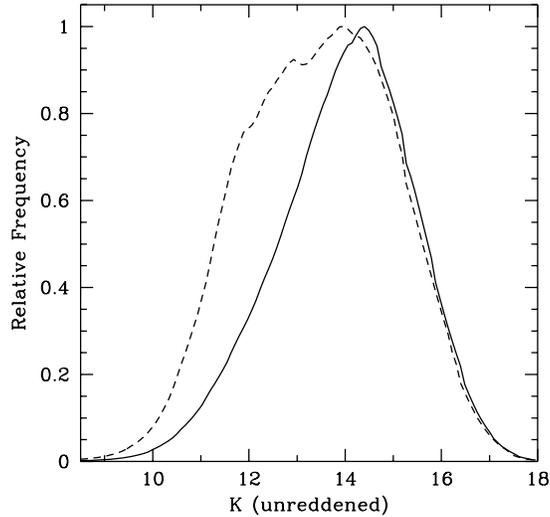}
  \caption{Distributions of intrinsic $K$ magnitudes for the mass donors in WNS XRBs for two WNS formation models (Pfahl \etal 2002).  The main difference between the two models is the relative proportion of binaries with intermediate- (solid line) and high-mass (dashed line) companions.}
\end{figure}

For the early-type donors of the WNSs, we would expect intrinsic
magnitudes of $K$=11-16, with the peak of the magnitude distribution
at $K\sim$14 (Figure 2); these are therefore readily distinguishable
from the majority of late-type donors expected for black hole X-ray
transients which generally have $K\geq$16 in quiescence.  The average
extinction towards the GC is $K\sim$3; therefore with our images,
which have a magnitude limit of $K$=20, we should detect most of the
WNSs.

\begin{figure}
  \includegraphics[height=.5\textheight]{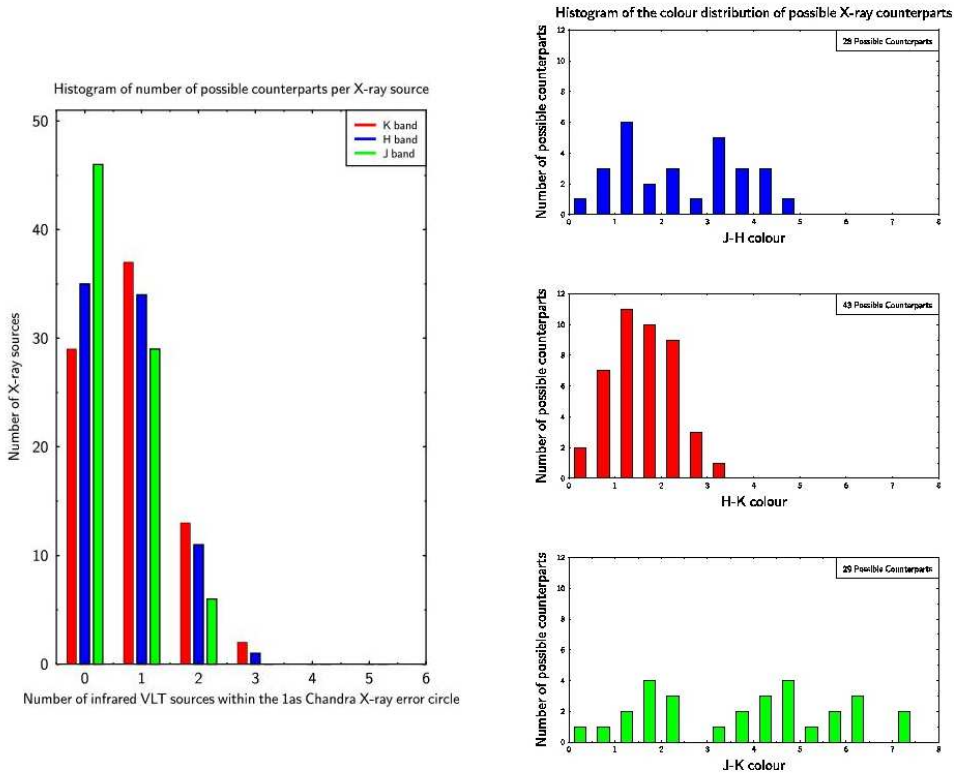}
  \caption{Left panel: histogram of the number of candidate IR counterparts per {\it Chandra} source within a 1'' X-ray error circle.  Right panel: distribution of IR colours of potential counterparts.}
\end{figure}

\section{Results: IR Magnitudes and Colours}

For 65\% of the X-ray sources in our VLT fields, there are one or two
resolved $K$-band sources within the 1'' {\it Chandra} error circle; only a
small number of X-ray sources have more than two potential
counterparts (Figure 3).  Over 50\% of the {\it Chandra} sources have no
potential $J$-band counterparts, and only a few of the potential IR
counterparts have colours consistent with unreddened foreground stars
(Figure 4).  This is consistent with the expectation that the majority
of the detected X-ray sources are heavily absorbed and thus are at or
beyond the GC.  

The magnitude and colour distribution of the identified candidate
counterparts is redder than expected for WNS systems (see Figures 3
and 4).  For an average GC extinction of $A_{K}\sim$2-3, the peak of
the expected reddened $K$ magnitude distribution for the WNSs is
$\sim$16-17.  The peak of the observed reddened $K$ magnitudes for the
potential counterparts is $\sim$14-15, with an {\it (H-K)} colour of
$\sim$1-2, as expected for later-type stars.  However, some potential
counterparts do have colours consistent with early-type stars.

\begin{figure}
  \includegraphics[width=.75\textwidth]{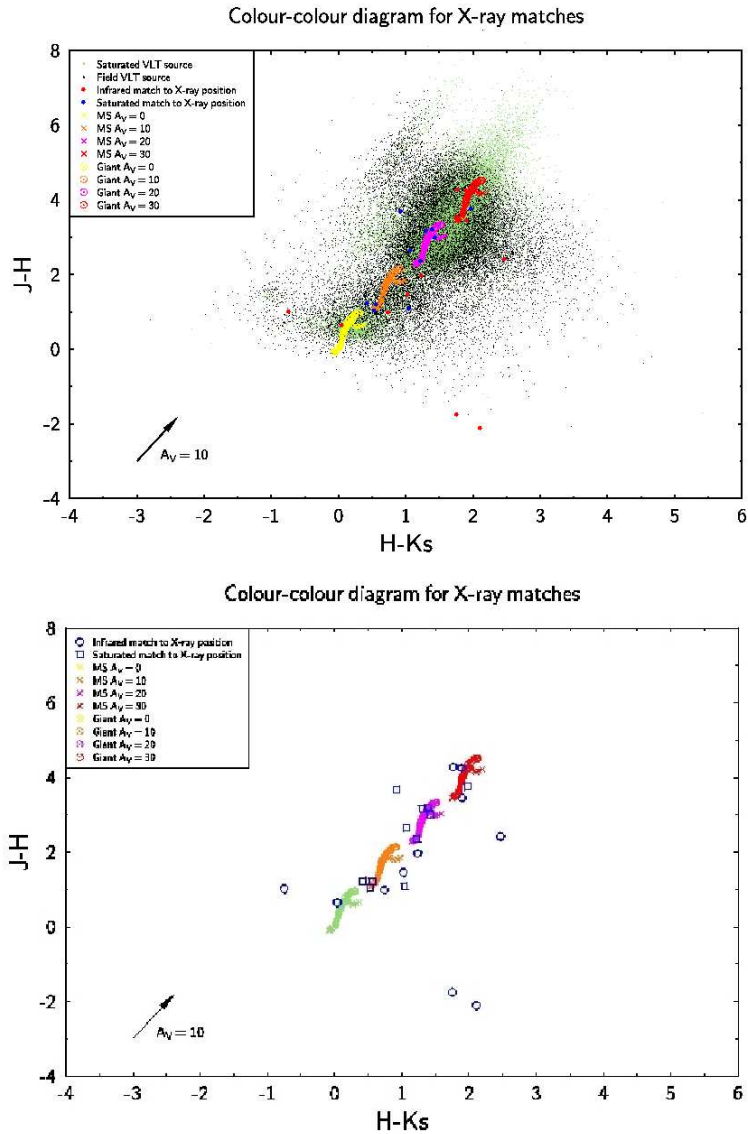}
  \caption{Top panel: colour-colour diagram showing all stars in our VLT fields. Bottom panel: colour-colour diagram of all potential IR counterparts to the {\it Chandra} sources for which we have full three-colour information.  The theoretical main sequence and giant branch are indicated at visual extinctions of 0, 10, 20, 30.  The IR colour-colour diagram illustrates that vast majority of field stars are consistent with highly reddened main sequence stars; thus most of the stars (including potential X-ray counterparts) are at the distance of the GC (or beyond).}
\end{figure}

There are no $K$-band counterparts for $\sim$35\% of the {\it Chandra}
sources.  This is larger than the expected fraction of background AGN
from the CDF estimate, though other groups have predicted larger
fractions (up to 50\%).  However, we note that the extinction in the GC
is extremely variable, even in the $K$-band.  This is evident from
visual inspection of our VLT data, which clearly show areas of heavier
than average extinction in the form of dust patches and lanes.
Therefore we need to carefully determine which X-ray sources actually
have no IR counterpart down to the $K$=20 magnitude limit, and which
are located in areas of locally heavy extinction.

\section{The Program Continues: IR Spectroscopy}

We have selected 36 of the best candidate counterparts for follow-up
IR spectroscopy (candidate magnitudes $K\sim$12-17); the goal of these
observations will be to identify the X-ray source counterparts via
detection of accretion signatures.  The primary accretion signature in
the $K$-band which distinguishes a true X-ray counterpart from a field
star is strong Brackett $\gamma$ emission; this technique of
identifying XRB counterparts has been verified with observations of
several well-studied GC XRBs (see \eg Bandyopadhyay
\etal 1999).  As these {\it Chandra} sources are weaker in X-rays than
the previously known population of Galactic XRBs, and thus have lower
accretion rates, the emission signature will likely be somewhat weaker
than in the more luminous XRB population.  However, the Brackett
$\gamma$ accretion signature is clearly detected in the IR spectra of
CVs, which are only weak X-ray emitters with a similar X-ray
luminosity range to the {\it Chandra} sources (see \eg Dhillon \etal
1997 for IR emission signatures in CVs).  Therefore we expect the
spectroscopic identification to be definitive even for these
low-luminosity X-ray sources.  For our brighter targets ($K$=12-14)
the spectra we obtain will allow us not only to identify the
counterpart via its emission signature but also to spectrally classify
the mass donors if absorption features are detected, a crucial step in
determining the nature of this new accreting binary population.  

Identifying IR counterparts to these newly discovered X-ray sources
provides a unique opportunity to obtain a census of the various
populations of accreting binaries in the GC and may ultimately allow a
determination of each system's physical properties.  As this {\it
Chandra} survey may contain 1\% of the entire population of accreting
binary systems in the Milky Way, our results will have important
implications for our understanding of XRBs in the Galaxy, including
their formation, evolutionary history, and physical characteristics.
The results of this observational program will represent an important
``calibration'' point for BPS codes so that they can be more reliably
applied to the study of other types of XRBs that have evolved from
massive stars, including ultraluminous X-ray sources (ULXs) in
external galaxies.  Finally, the combination of the imaging data and
our spectroscopic follow-up will allow us to identify the nature of an
entirely new population of X-ray emitters within our Galaxy.






\bibliographystyle{aipprocl} 


\IfFileExists{\jobname.bbl}{}
 {\typeout{}
  \typeout{******************************************}
  \typeout{** Please run "bibtex \jobname" to optain}
  \typeout{** the bibliography and then re-run LaTeX}
  \typeout{** twice to fix the references!}
  \typeout{******************************************}
  \typeout{}
 }



\end{document}